# Electronic structure of Pt-doped superconductor $CaFe_{1-x}Pt_xAs_2$

Wasim Raja Mondal[1] and Swapan K. pati[1]

Theoretical Sciences Unit[1]

Jawaharlal Nehru Centre for Advanced Scientific Research, Bangalore, 560064, India.

email addresse:

wasimr.mondal@gmail.com

We have computed complete electronic structure of new Pt-doped superconductor $CaFe_{1-x}Pt_xAs_2$. Our findings for the parent compound of this superconductor agree well with experimental predictions.

# Introduction

Since the discovery of copper oxide (YBaCuO), high temperature superconductors (HTS) with transition temperature ($T_C$) 93 K in 1987 [1], there have been innumerable research in finding cuprates with higher and higher $T_C$. Up to year 2009, the highest-temperature superconductor found is cuprate-perovskite ($HgBa_2Ca_{m-1}Cu_mO_{2m+2+\Delta}$) [2] with $T_C$ =164 K, under quasi hydrostatic pressure. At the same time, there have been efforts to synthesize similar type of superconductors other than cuprates. Some examples are $MgB_2$ ($T_C$ =39K) [3], $Cs_3C_{60}$ ($T_C$=38K) [4]. In the meantime, prediction of superconductivity in iron-pnictide ($La[O_{1-x}F_x]FeAs$, $T_C$=26K) has attracted much attention [5]. In particular, the transition temperature ($T_C$) in $SmO_{1-x}F_xFeAs$ is found to be 55 K [6]. Such high transition temperature like that of cuprates is very unconventional and becomes more challenging as Bardeen, Cooper, Schrieffer (BCS) theory [7] cannot explain the microscopic mechanism behind high-temperature superconductivity. Again, these iron-pnictides are becoming very significant because of their similarities with cuprates, which can provide us appropriate starting point to predict superconducting mechanism in cuprates.

Recently, an iron-pnictide superconductor, $CaFe_{1-x}Pt_xAs_2$, has been synthesized by Kawamata et. al. [8]. Two types of structures have been reported for this superconductors, $SrZnBi_2$ (type-A) and $HfCuSi_2$ (type-B). The space group of type-A is I4/mmm and that of type-B is p4/nmm. Type-A structure has two $Fe_{1-x}Pt_xAs$ layers whereas type-B has only one layer. Actually, there is difference in the stacking of layers along the c-axis. (Fig. 1 (a), 2 (b) ). It has been found that there is small difference in the transition temperature ($T_c$) of these two structures; where type-A has $T_c$ =29.3 K and for type-B it is $T_c$=30.1 K. The novelty of this compound is that, it acts as a HTS in the presence of large fraction of dopants like platinum and it does not have $s_{+-}$ order parameter symmetry. But surprisingly, CaFe2As2 is not superconducting in the presence of Pt dopants [9]. That is why; $CaPt_xFe_{1-x}As_2$ is interesting and demands theoretical understanding for its observed behavior. In this work, we study both type-A and type-B structures. We have computed the electronic structure and its properties using density functional theory. In this paper, we mainly concentrate on the following questions in order to understand the physics of this system. (1) What kind of magnetic ordering is present in the parent compound of this superconductor? (2) Besides the dominant contribution of Fe '$3d$' orbitals, what is the role of doped Pt atom in the electronic structure of the system? (3) What is the role of topology of Fermi surface in the pairing mechanism? (4) What is the nature of the bonding between Fe and As? (5) What is the symmetry of the superconducting order parameter?

## Computational Detail

Estimated effective U in case of iron-pnictides is assumed to be less than 0.5 eV. This value is reduced by hybridization of localized Fe '$3d$' electrons with As '$4p$' electrons [10], and hence we can apply band theory for iron-pnictides. Our results are based on first-principles pseudo potential-based density functional theory (DFT) as implemented in the PWSCF package [11]. We use ultra-soft pseudo potentials [12] to describe the interaction between ion core and outer electrons and plane-wave basis set with kinetic energy cut-off 300 Ry for charge density and 30 Ry for wave functions. We consider generalized gradient approximation using Perdew-Burkew-Enzerhoff parameterized for the exchange correlation function. For spin ordered phases, we perform spin-polarized calculations with different initial spin configurations. For type-A, we consider unit cell but for type-B we consider 1×1×2 super-cell for comparison with type-A. We take 9×9×3 Monkhorst-pack mesh [13] for the integration over the Brillouin zone for both types of structures. Optimization is done by energy minimization in the Broyden-Fletcher-Goldfarb shanno(BFGS) [14]-based method. For the metallic behavior of the system, we use Methfessel-paxton scheme [15] with a smearing width of 0.003 Ryd for the occupation numbers. Structural Optimization is carried out with total energy difference between last two consecutive steps less than $10^{-8}$ Ryd and the maximum force on each atom less than 0.002 Ry/ Bohr.

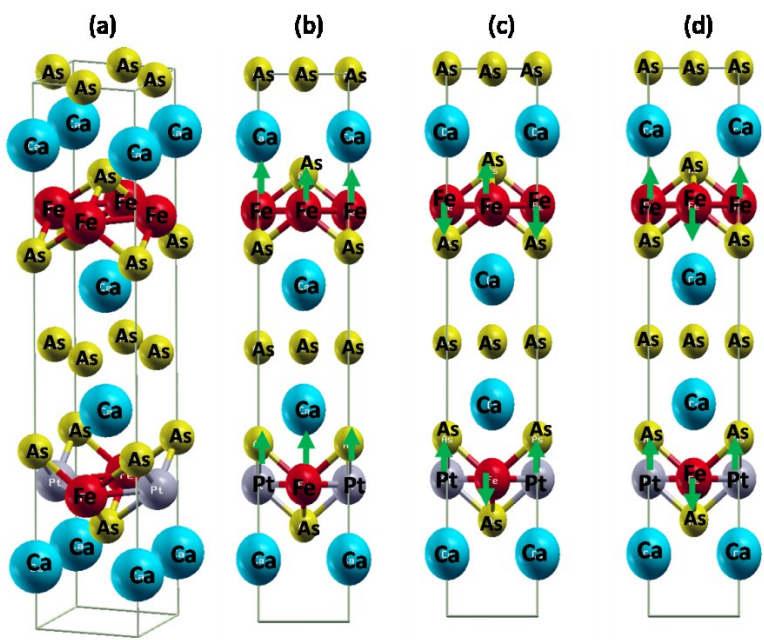

**Figure 1: (a) crystal structure of type-A configuration. The schematic diagrams of different spin ordering on the Fe atoms in the unit cell, (b)Ferromagnetic ordering, (c)antiferromagnetic (AFM1) ordering, (d)antiferromagnetic stripe (AFM2) ordering.**

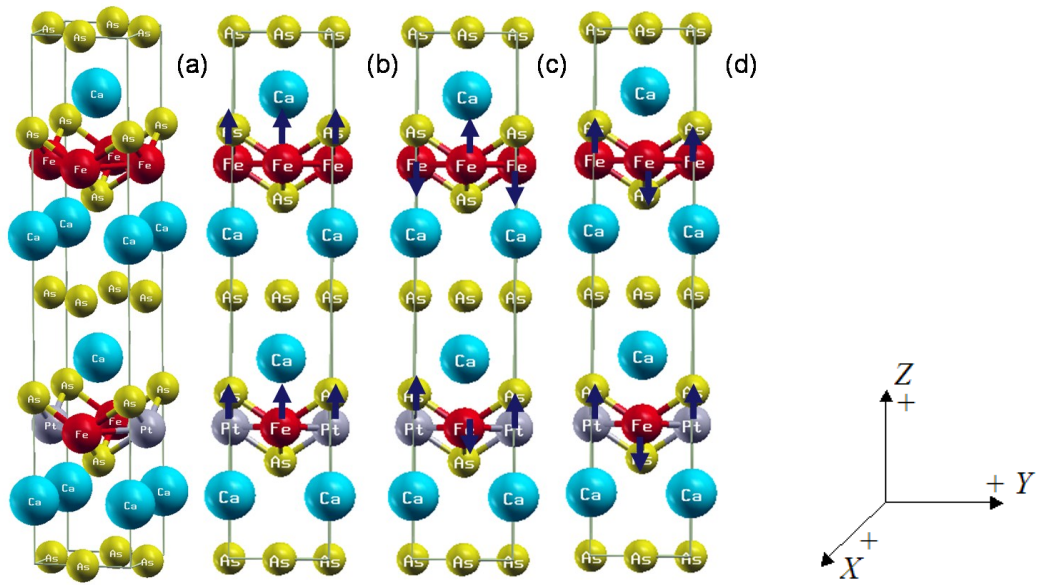

**Figure 2:** (a)Crystal structure of type-B configuration. The schematic diagrams of different spin ordering on the Fe atoms in 1×1×2 supercell. (b)Ferromagnetic (FM) ordering (c) Antiferromagnetic (AFM1) ordering (d) Antiferromagnetic stripe (AFM2) ordering.

**Table 1.** Energy value of different spin-configurations and absolute magnetic moments (|M| in $\mu_B$ /atom) of the system (type-A). Total energies (eV) are given for all the spin configurations with respect to the total energy of non-mgnetic (NM) state.

| States | Energy(eV) | Absolute Magnetic Moment $|M|(\mu_B)$ |
|---|---|---|
| AFM2 | -0.213 | 1.947 |
| AFM1 | -0.210 | 1.945 |
| FM | -0.044 | 0.940 |
| NM | 0.000 | 0.000 |

**Table 2.** Optimized lattice parameters of the Pt-doped crystal, internal structure parameters ($Z_{CaI}$ and $Z_{CaII}$), and bond-length between Fe-Fe of the system (type-A). Here $Z_{CaI}$ and $Z_{CaII}$ are the height of Ca atom from above and below the Fe-layer respectively.

| States | Lattice parameter (Å) | | $Z_{ca-I}$ | $Z_{Ca-II}$ | Bond-length(Å) (Fe-Fe) |
|---|---|---|---|---|---|
| | a | c | | | |
| AFM2 | 3.971 | 20.338 | 0.1365 | 0.1355 | 2.80 |
| AFM1 | 3.974 | 20.295 | 0.1364 | 0.1355 | 2.81 |
| FM | 3.951 | 20.281 | 0.1383 | 0.1306 | 2.79 |
| NM | 3.935 | 20.169 | 0.1376 | 0.1315 | 2.78 |
| EXPT.[8] | 3.920 | 20.716 | 0.1305 | 0.1305 | 2.77 |

**Table 3.** Energy value of different spin-configurations and absolute magnetic moments (|M| in $\mu_B$ /atom) of the system (type-B). Total energies are given for all the spin configurations with respect to the total energy of the non-magnetic(NM) state.

| states | Energy (eV) | Absoulete magnetic moment (|M|($\mu_B$) |
|---|---|---|
| AFM2 | -0.216 | 1.94 |
| AFM1 | -0.149 | 1.16 |
| FM | -0.04 | 2.19 |
| NM | 0.00 | 0.00 |

**Table 4.** Optimized lattice parameters of the Pt-doped crystal, internal structure parameters ($Z_{CaI}$ and $Z_{CaII}$) and bond-length between Fe-Fe of the system (for type-B).

Here $Z_{CaI}$ and $Z_{CaII}$ are the height of Ca atom from above and below the Fe-layer respectively.

| states | Lattice parameter a (Å) | Lattice parameter c(Å) | $Z_{CaI}$ | $Z_{CaII}$ | Bond(Å) (Fe-Fe) |
|---|---|---|---|---|---|
| AFM2 | 3.965 | 20.385 | **0.135** | 0.134 | 2.80 |
| AFM1 | 3.945 | 20.328 | 0.135 | 0.134 | 2.79 |
| FM | 3.92 | 20.81 | 0.135 | 0.134 | 2.77 |
| NM | 3.933 | 20.208 | 0.137 | 0.130 | 2.78 |
| EXPT.[8] | 3.902 | 21.024 | 0.130 | 0.130 | 2.77 |

# Results and Discussion

The structure is optimized with respect to both atomic positions and experimental lattice constants. To investigate the role of magnetic ordering on the electronic properties, we consider non-magnetic and different magnetically ordered phases depending on the spins on Fe atoms. In the ferromagnetic phase (FM), all spins on Fe atoms are in same direction in both y and z directions, whereas, in the G-type anti-ferromagnetic (AFM1) phase, these spin-ordering is anti-ferromagnetic in y-direction but ferromagnetic in z-direction. In the case of anti-ferromagnetic stripe (AFM2) phase, spin-ordering is anti-ferromagnetic in both y and z-direction. All these spin configurations are given in Fig. 1 and Fig. 2.

Our estimated energies of the different spin configurations for type-A are given in Table1. We find that AFM2 phase is energetically more stable with respect to all the other spin-configurations. The anti-ferromagnetic stripe (AFM2) phase is energetically more stable than the nonmagnetic phase (NM) by -0.213 eV/unit cell but the energy difference between anti-ferromagnetic striped (AFM2) and G-type anti-ferromagnetic (AFM1) phase is 3 meV/unit cells. Hence, G-type anti-ferromagnetic (AFM1) phase can be one of the meta-stable states of the system. But the large difference in energy of AFM2 phase and NM phase clearly indicates the importance of magnetic ordering in the structural stability of the system. Our calculated lattice parameters are presented in Table 2. In the case of AFM2 phase, calculated value of lattice parameter 'a' is 1.28% larger than the experimental value and lattice parameter 'c' is 1.8% smaller than the experimental value. From Table 2, it is clear that there is significant difference between the calculated and experimental $Z_{CaI}$ and $Z_{caII}$ which is the height of pnictogen atom (Ca) from iron-arsenide plane. In the case of the Type-B system, we again find the AFM2 phase to be energetically more stable than any other spin configurations of this system as shown in Table 3. The calculated lattice parameter 'a' is 1.5% larger and lattice parameter 'c' is 3.1% smaller than the experimental value [8]. Like type-A system, there is also a significant difference in the separation of Ca atoms from Fe-plane.

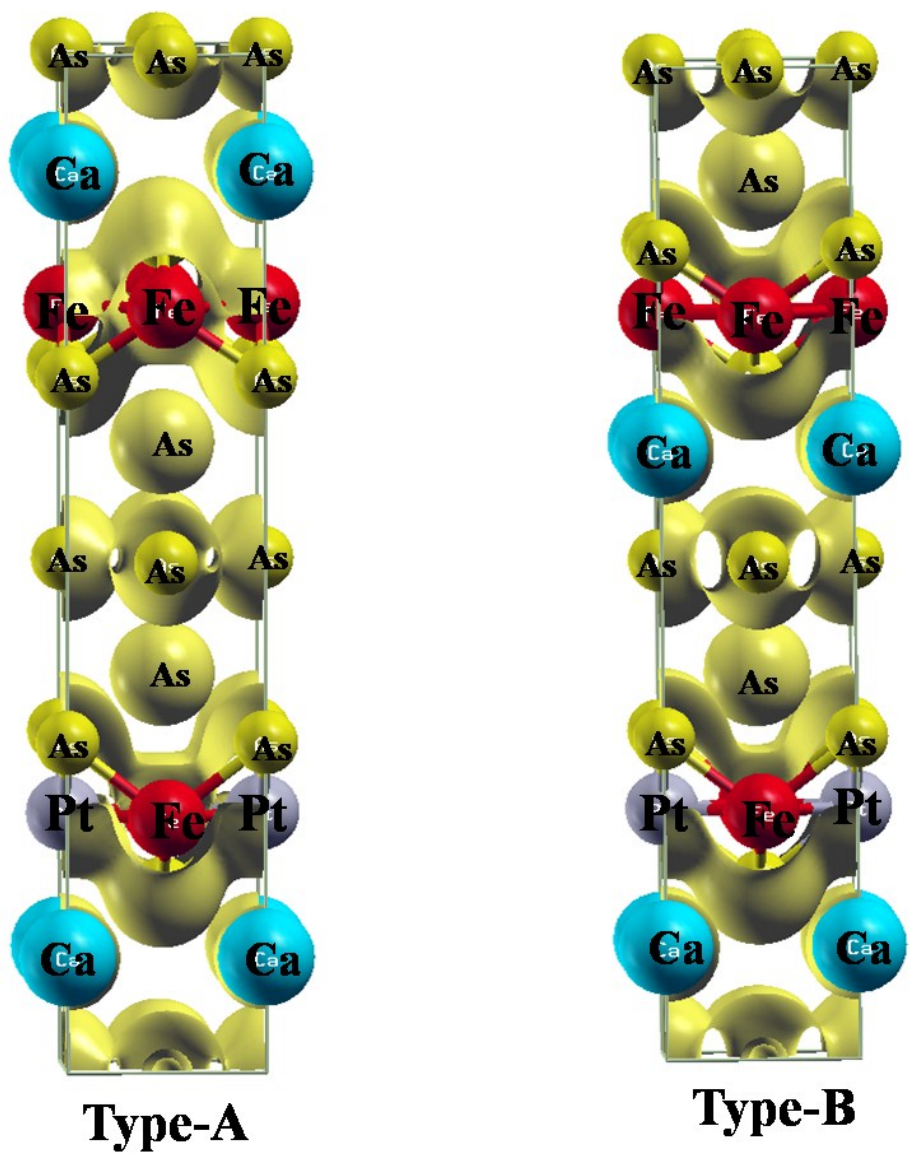

**Figure 2: Charge density plot for Type-A and Type-B configurations.**

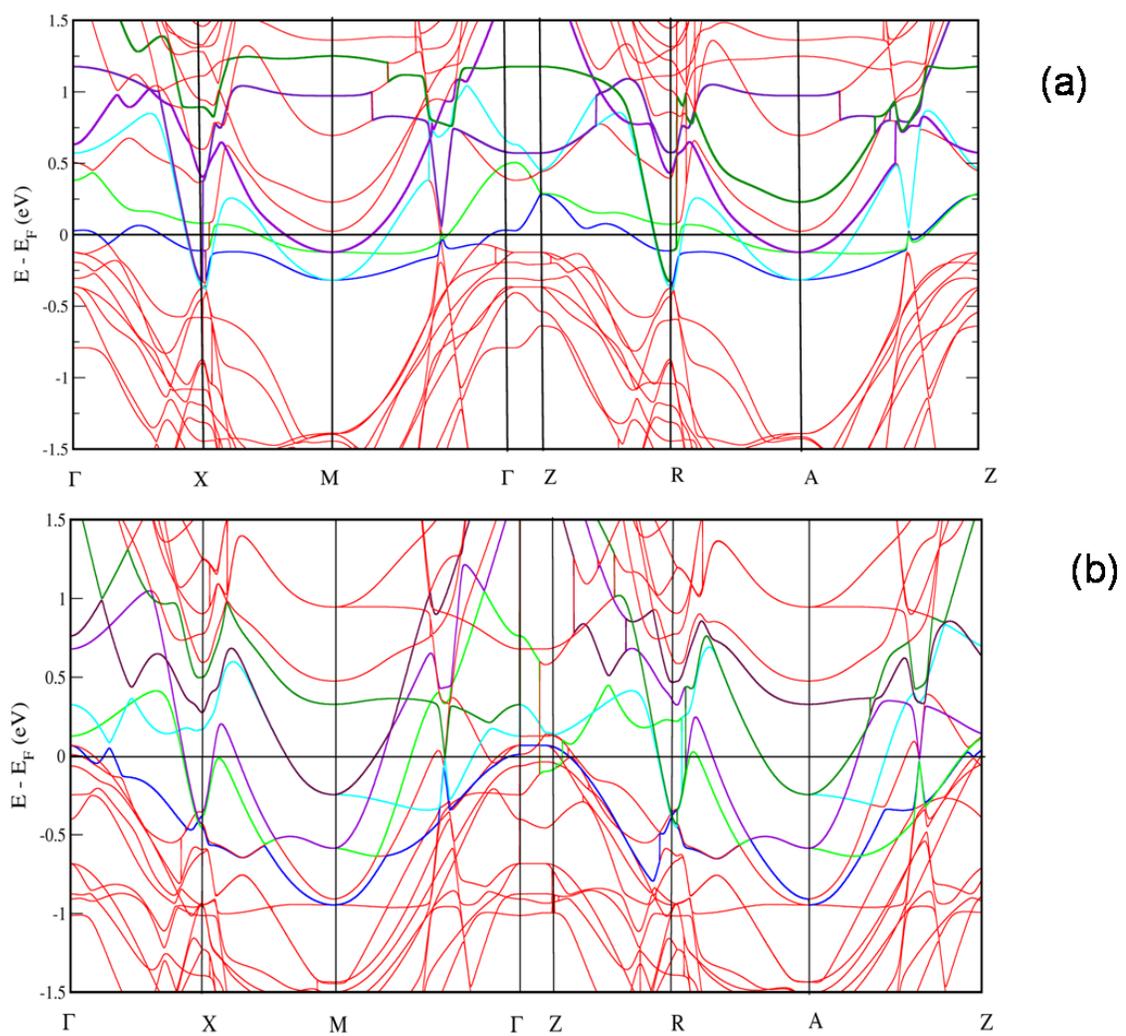

**Figure 3: Electronic band structure of type-A: (a) NM phase, (b) AFM2 phase.**

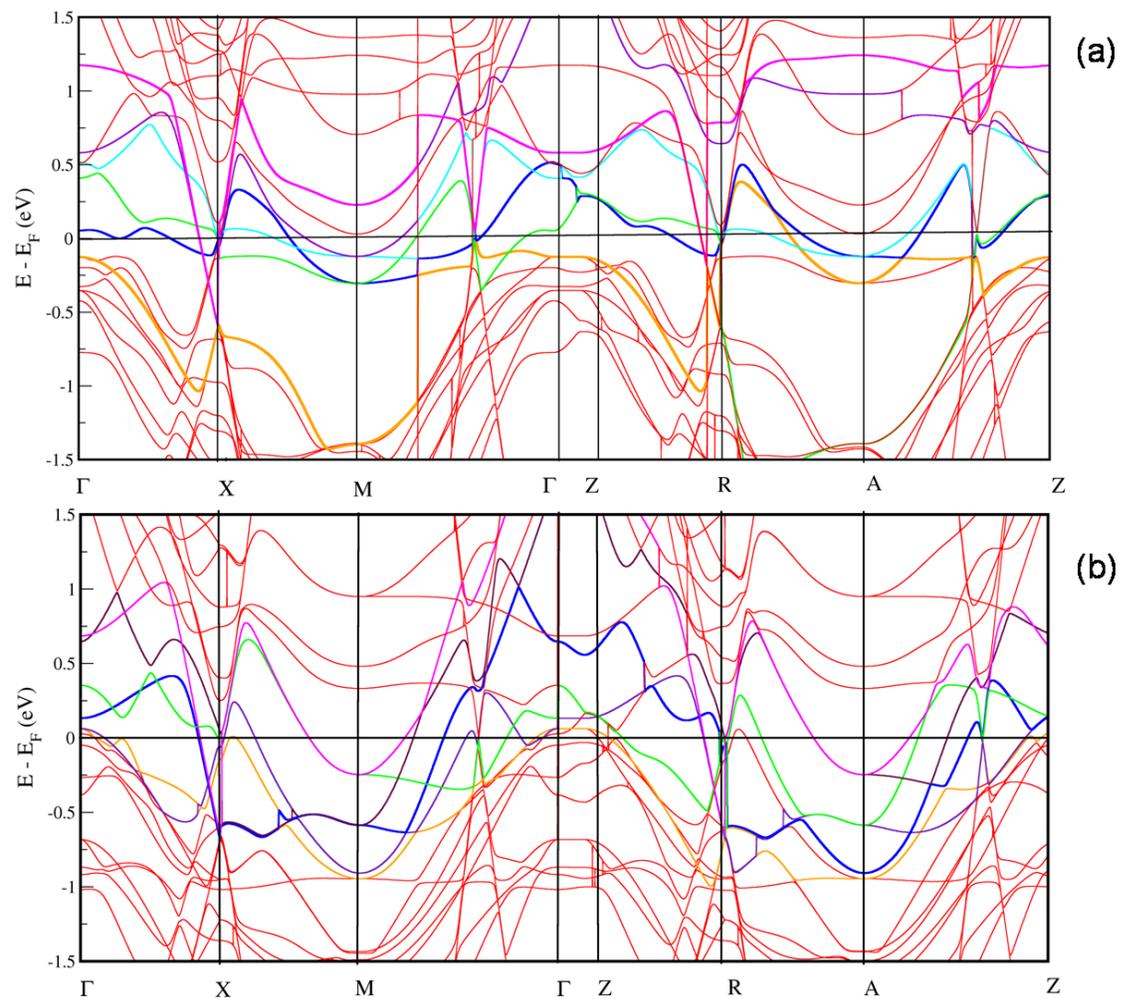

**Figure 4: Electronic band structure of the type-B: (a) NM phase, (b) AFM2 phase.**

Hence in both the structures, we find that the lattice parameters show anisotropic deformation of the crystal structure due to presence of magnetic ordering. For both Type-A and Type-B, in the optimized structures with different magnetic ordering, there exists a variation of pnictogen height where a small variation of pnictogen height gives a drastic change in the electronic structure as reported in Refs. [16]. AFM2 stripe phase is an important state of the parent compound of this high temperature superconductor. From data of Table 1 and Table 3, it is interesting to note that magnetic moment in AFM2 stripe phase is maximum compared to all other spin configuration in both type-A and type-B structure. In addition, lattice distortion in AFM2 stripe phase is also maximum along z direction. That means, lattice parameter along z direction increases with the increase in magnetic moment value. To understand this behavior we have plotted charge distribution for both the configuration in Fig. 3. There is a significant overlap between the electron cloud of Fe '3d' and As '4p' orbital (see Fig. 3). Besides this, there is also overlap between As 4p electrons between each other. This tells us the hybridization character between Fe 3d with As 4p electrons. Change in lattice parameter 'c' can be understood under the following way as described by Yildirim [17]:

Pnictogen atom such as As situated symmetrically top and below the iron-plane. There is tendency of overlap of As ions above the Fe-plane with the As ions below the same iron-plane. There are also interactions between Fe and As. This Fe-As interaction increases due to increase in Fe-moment. This increase in Fe-As interaction prevents to overlap of As ions above the iron-plane with As ions below the iron-plane. That is why; we observe the increase of lattice parameter c with the increase in Fe-moment. But it is interesting to note that, our calculated value of c parameter is not that much deviated from experimental value as predicted by Yildrim in $CaFe_2As_2$ system. In our system, we have substituted Fe by Pt partially. This substitution decreases the magnetic moment. Hence, we have not observed that much strong increase in lattice parameter c value as found in $CaFe_2As_2$

Our calculated band structures for $CaFe_{1-x}Pt_xAs_2$ system with both I4/mmm symmetry and p4/nmm symmetry at the equilibrium lattice parameters along various high-symmetry points are given in Fig. 4 and Fig. 5. To understand the role of magnetism, we have investigated the band structures for both NM and AFM2 states of the systems. The overall band structures of both type-A and type-B configurations show the same features. If we consider the band structute of NM and AFM2 phases, they are almost same but there is splitting of the degenerate energy states of NM phase in AFM2 phase. This splitting of energy levels is due to the broken symmetry of the AFM2 striped phase because in this phase spin ordering is antiferromagnetic along y direction but ferromagnetic long z direction. In the band diagrams, bands are highly dispersive and we do not observe any flat-bands near the Fermi level in both (AFM 2 and NM) configurations. Since electrons near the Fermi-level play a crucial role in the formation of the superconducting state, we analyze the electronic structure near the Fermi level. Bands near the Fermi level (in the range from -1.5 eV to +1.5 eV ) comes mainly from Fe 3d states with small contribution of Pt '1d' and As '*2p*' states. We consider this system metallic rather than semi metallic in the sense that there are six bands crossing the Fermi level along the Γ-Z line. These bands with energies of 0.03eV, 0.39eV, 0.57eV, 0.65eV and 2.03eV at the Γ- point as shown in Fig. 4 and Fig. 5 Most of them cross the Fermi level in the directions Γ-X, Γ-M, R-A, Z-A in the Brillion zone and create the complex shape of the Fermi surface (see Fig. 7). The bands that intersect the Fermi-level along the X-Γ direction, form electron-like sheet of the Fermi surface and bands cross the Fermi-level in the direction of M-R create hole-like sheet of the Fermi surface. Hence in the Fermi surface, we observe small hole-pocket at the Z-point and electron-pocket at the M point (see Fig. 6 and Fig. 7).

To explore the possible order-parameter for this system, we observe the band structure and Fermi-surface more carefully. Generally, $d_{x^2-y^2}$ order parameter will be most effective when the majority of states are found near $(\pi, 0, k_z)$ and equivalent points, while the $s_{+-}$ order parameter will be considered when the majority of states are found near the $(0,0,k_z)$ and $(\pi,\pi,k_z)$ points [18]. In most of the iron-pnictide study, we observe the combination of hole pocket at the Γ point, and an electron pocket at the X point of the unfolded Brillion Zone. This is not the case in $CaFe_{1-x}Pt_xAs_2$. Our calculation shows that there are significantly no states available for superconductivity near Γ-point and no large cylindrical Fermi surface at $(\pi, \pi, k_z)$. This predicted result is similar with very recent report on Ce- and Pu- based superconductor (115 family) [18]. We find small hole pocket centered at the Z point and electron pocket centered at M point (see Fig 2.6 and Fig 2.8). Another interesting feature of this Fermi surface is that it has low dispersion along $k_z$ direction. Hence our finding excludes the possibility of $s_{+-}$ symmetry of the superconducting order parameter. This observation is consistent with the experimental prediction of the symmetry of the superconducting order parameter of this system [8].

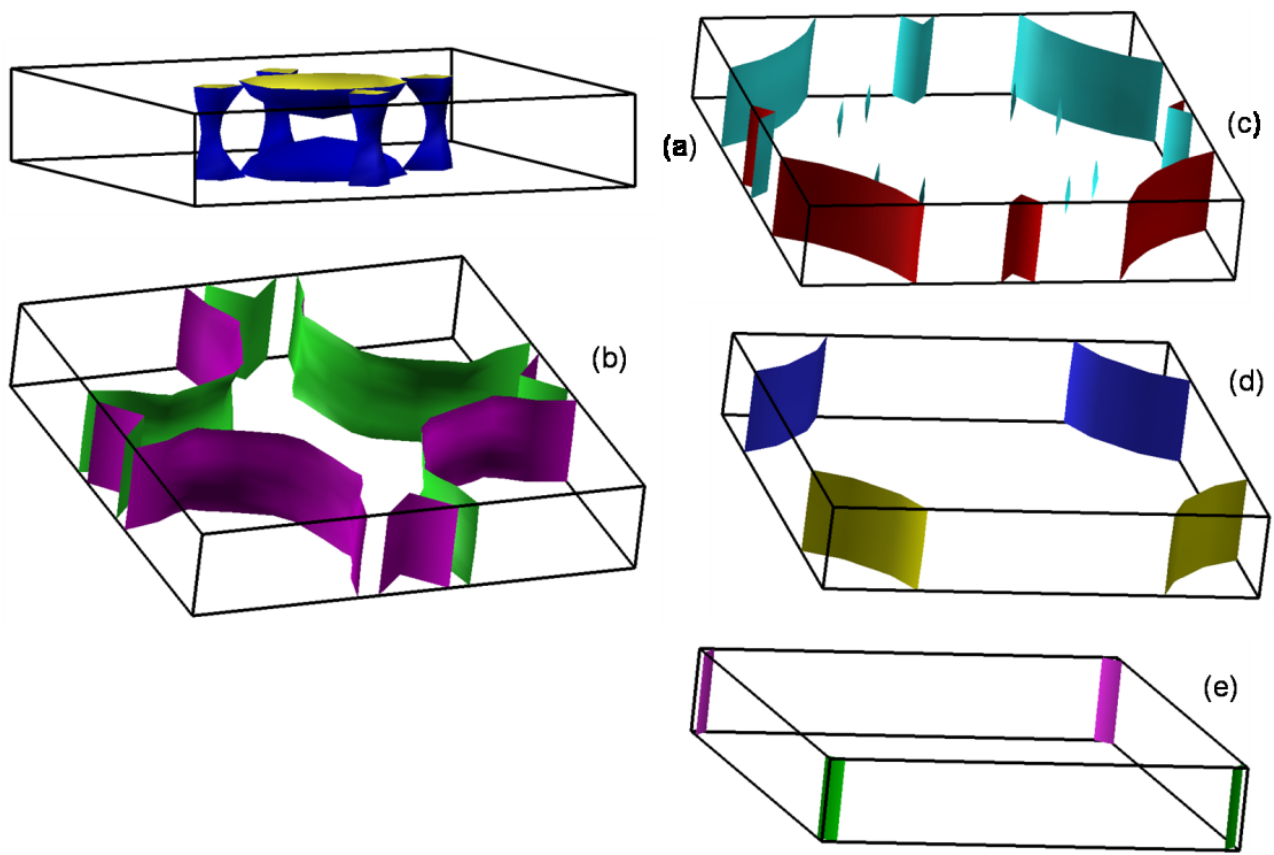

**Figure 5: Separate sheets of the Fermi surface for type-A configuration. Four of them (b,c,d,e) are electroniclike, and sheet (a) is holelike**

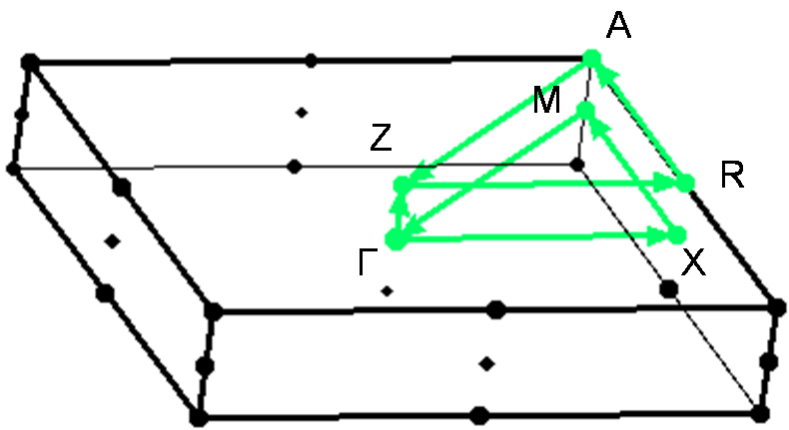

**Figure 6: Diagram of path taken in brillouin zone along different symmetry point.**

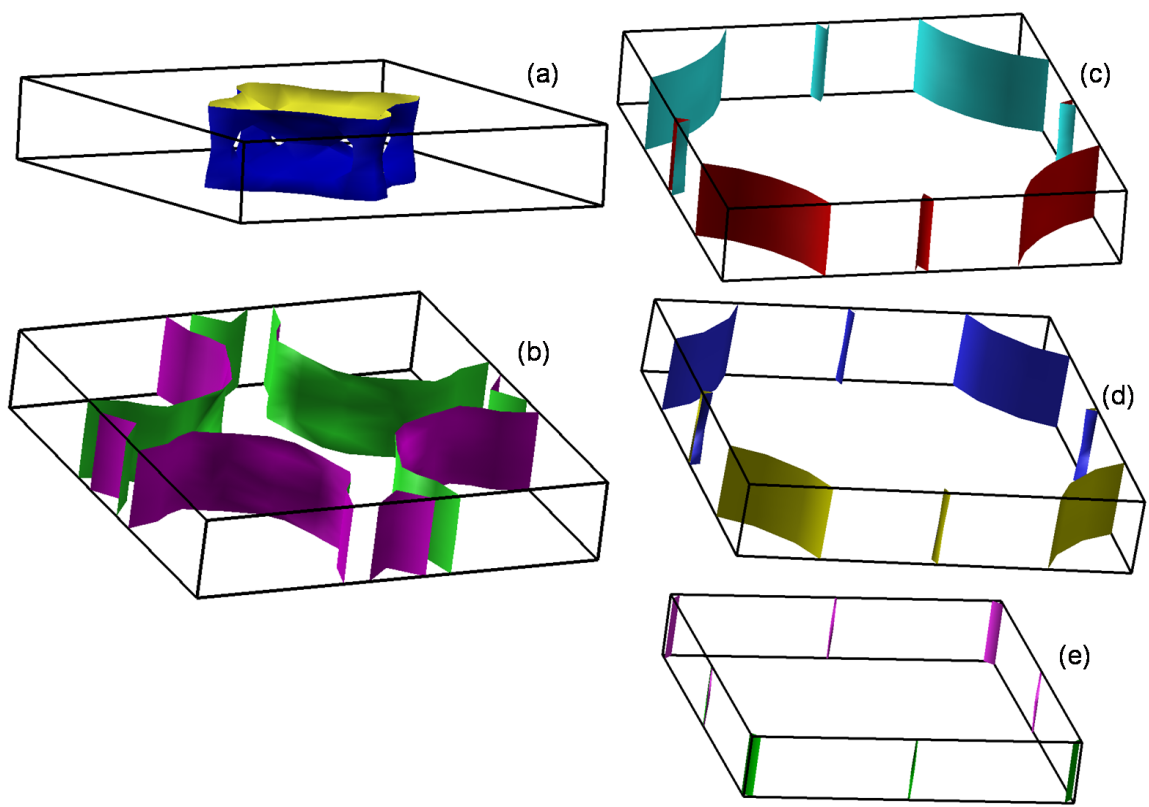

**Figure 7:** seperate sheets of the Fermi surface for type-B configuration. Four of them (b,c,d,e) are electron-like and sheet (a) is hole like

From the density of state plots of both the states (AFM2 and NM) in figure 9 and figure 10, we find that there is finite density of states in the Fermi level which shows the metallic character and agrees with experimental findings. The difference in the electronic structure of AFM2 and NM phase will be clear if we compare the density of states of these two phases. At the Fermi energy for the NM phase the density of state $N(E_F)=11.21$ and for AFM phase $N(E_F)=6.37$ (see Fig. 9 and Fig. 10). From the orbital density of states which is shown in Fig. 2.11, we find that contribution of Fe '*3d*' orbital to the density of states $N(E_F)$ is dominant at the Fermi level . Clearly, this is due to the effect of direct intra-layer Fe-Fe interactions. This feature is consistent with the other layer-type tetragonal Fe-based superconductors.[19-20] The bandwidth of the Fe band around the Fermi level is about 4eV (-2 eV to +2 eV). In this energy ranges, besides the dominant contribution of Fe '*3d*' orbital, there is a small contribution of As '4p' and Pt '5d' orbitals too. We find that Fe 3d states overlap with As 4p states at -2 eV. This p-d hybridization makes Fe-As bonding strongly covalent. There is also a significant contribution from Pt '5d' orbital because of high concentration Pt doping. However, there is negligible contribution of Ca atoms to the total density of states at the Fermi-level. So Fe-layers containing Fe '*3d*' orbitals play the key role in the electronic structure of the system. In Fig. 2.12, we have plotted also the contribution of different Fe '*3d*' orbital for both type-A and type-B systems. Both the systems show a common feature. For both the systems, the spin-up and spin-down components of Fe atoms are partially filled. This kind of occupation of Fe '*3d*' state reveals the origin of the calculated local magnetic moment 1.94 $\mu_B$ of Fe atom in the AFM2 stripe phase for both Type-A and Type-B structure.

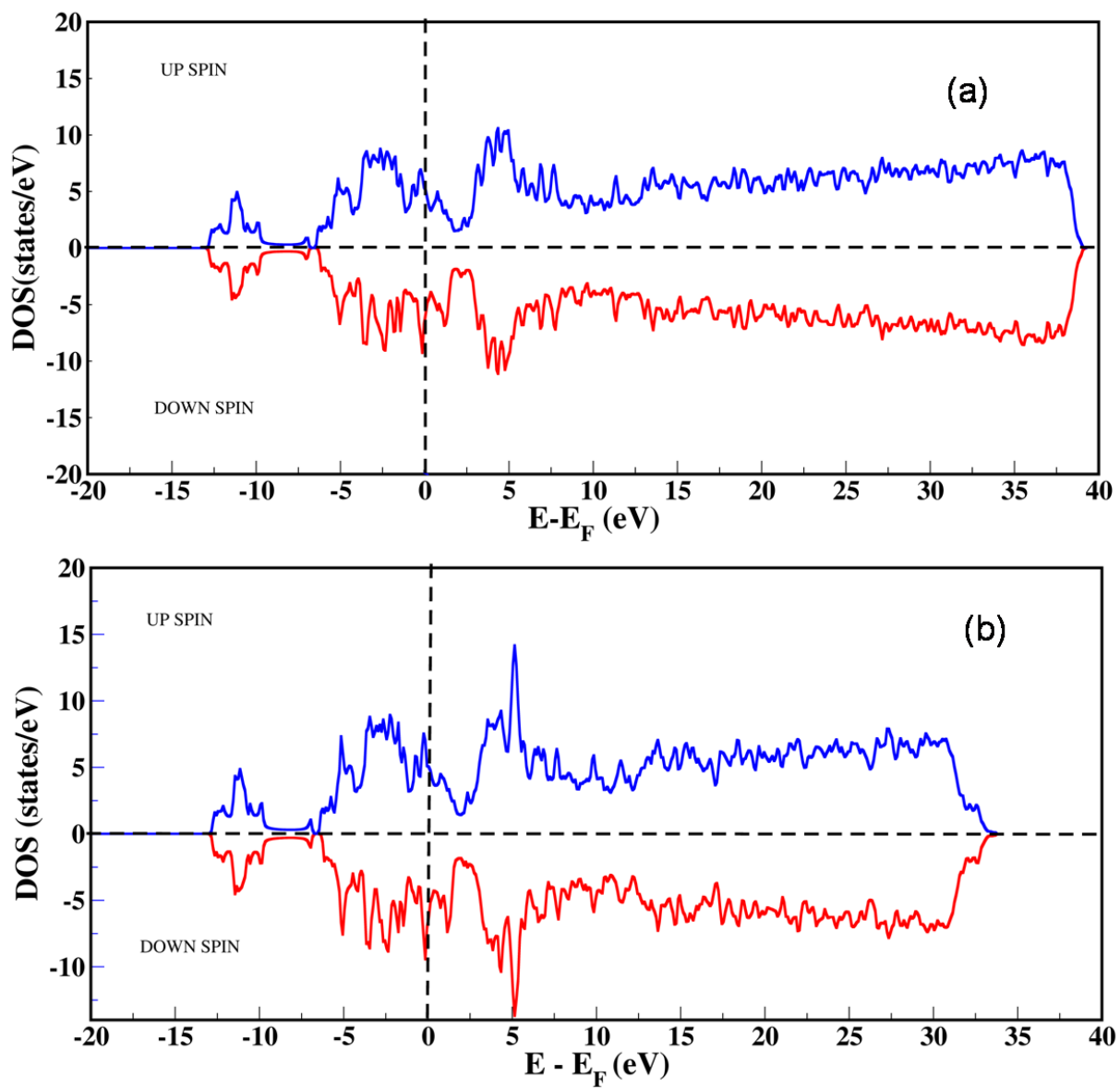

**Figure 8:** Density of states plot for AFM2 stripe-phase: (a) type-A (b) type-B

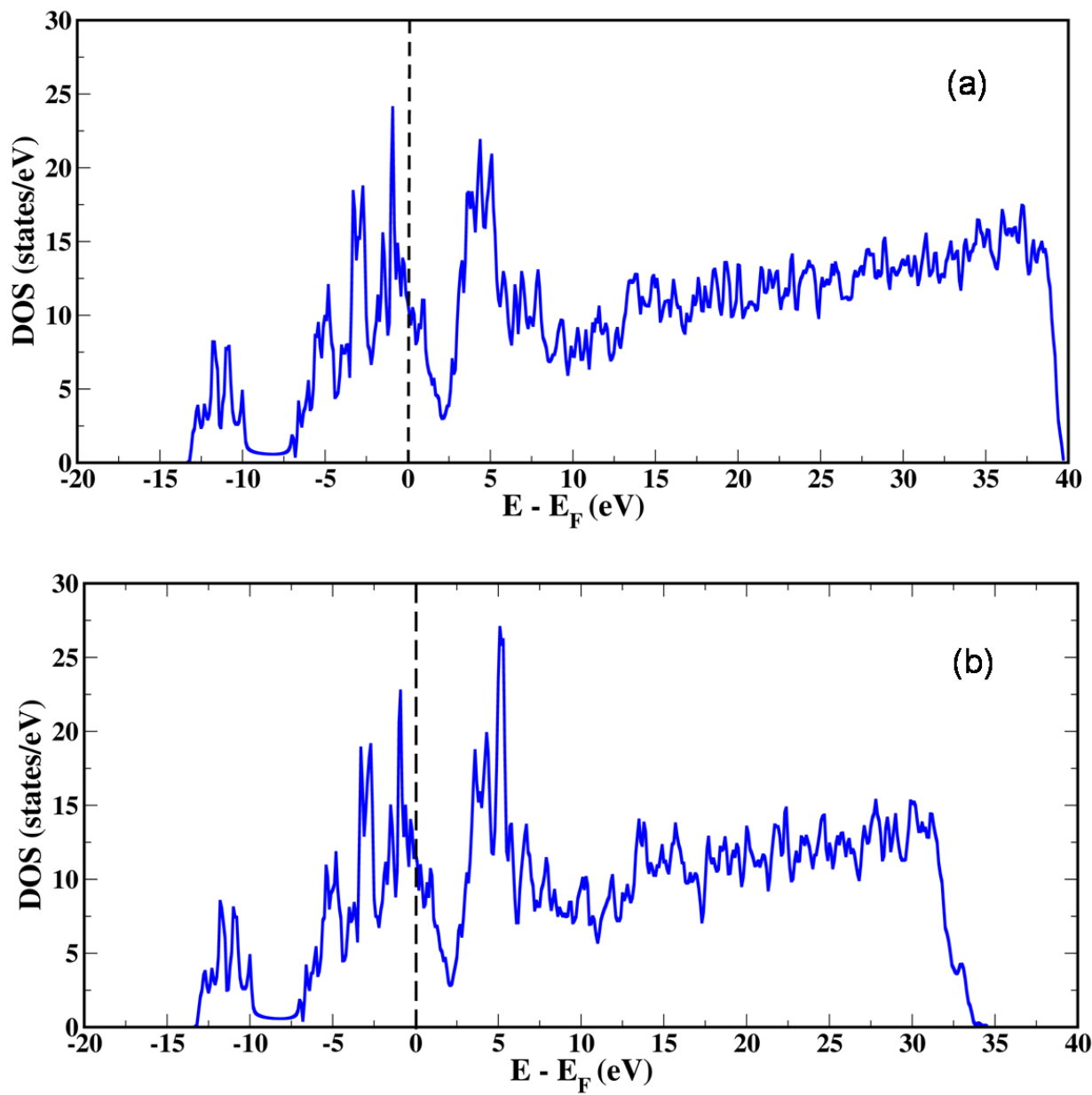

**Figure 9: Density of states for non-magnetic (NM) phase: (a) type-A (b) type-B**

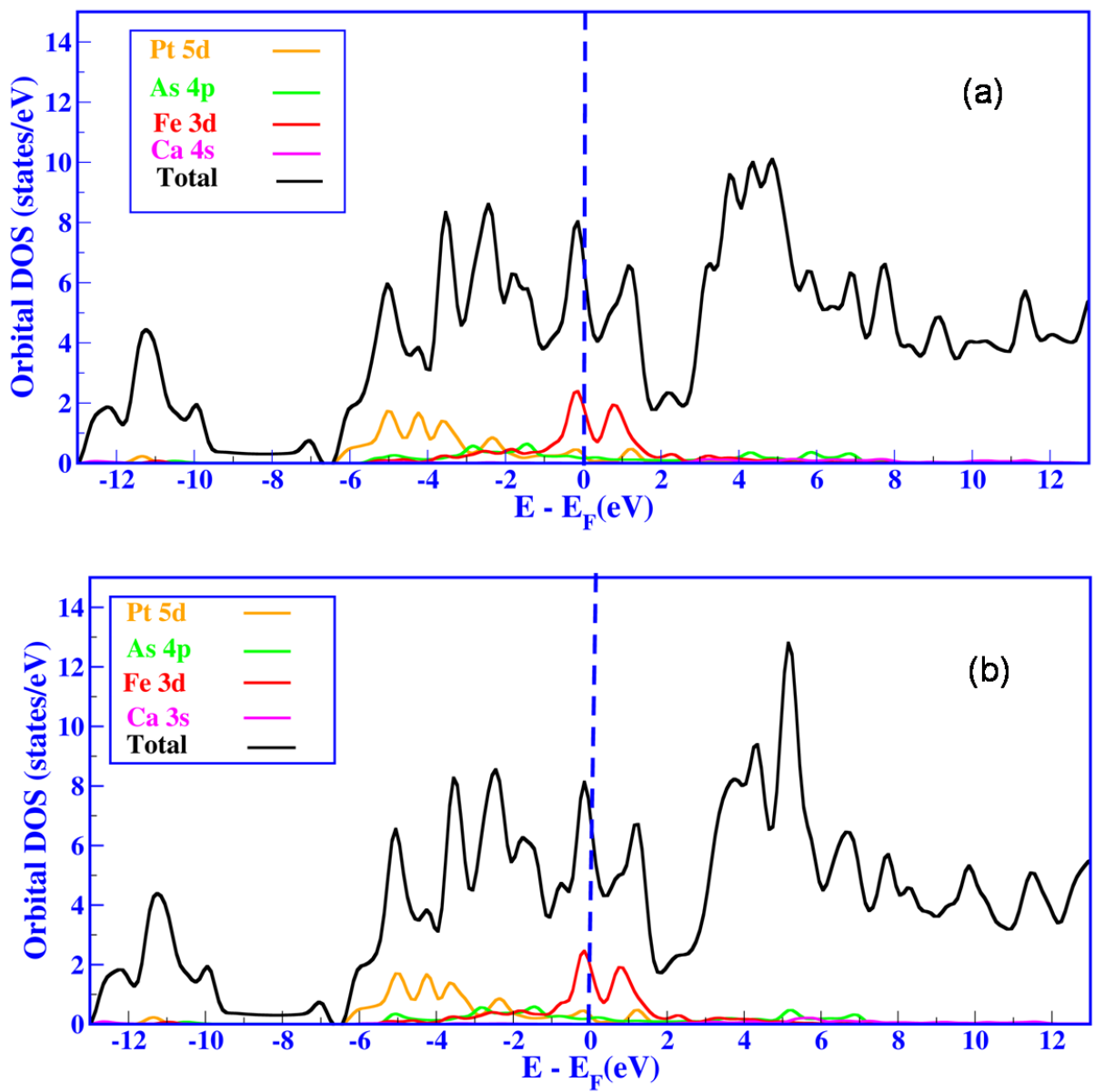

**Figure 10: Orbital density of states for AFM2 phase (a) type-A (b) type-B**

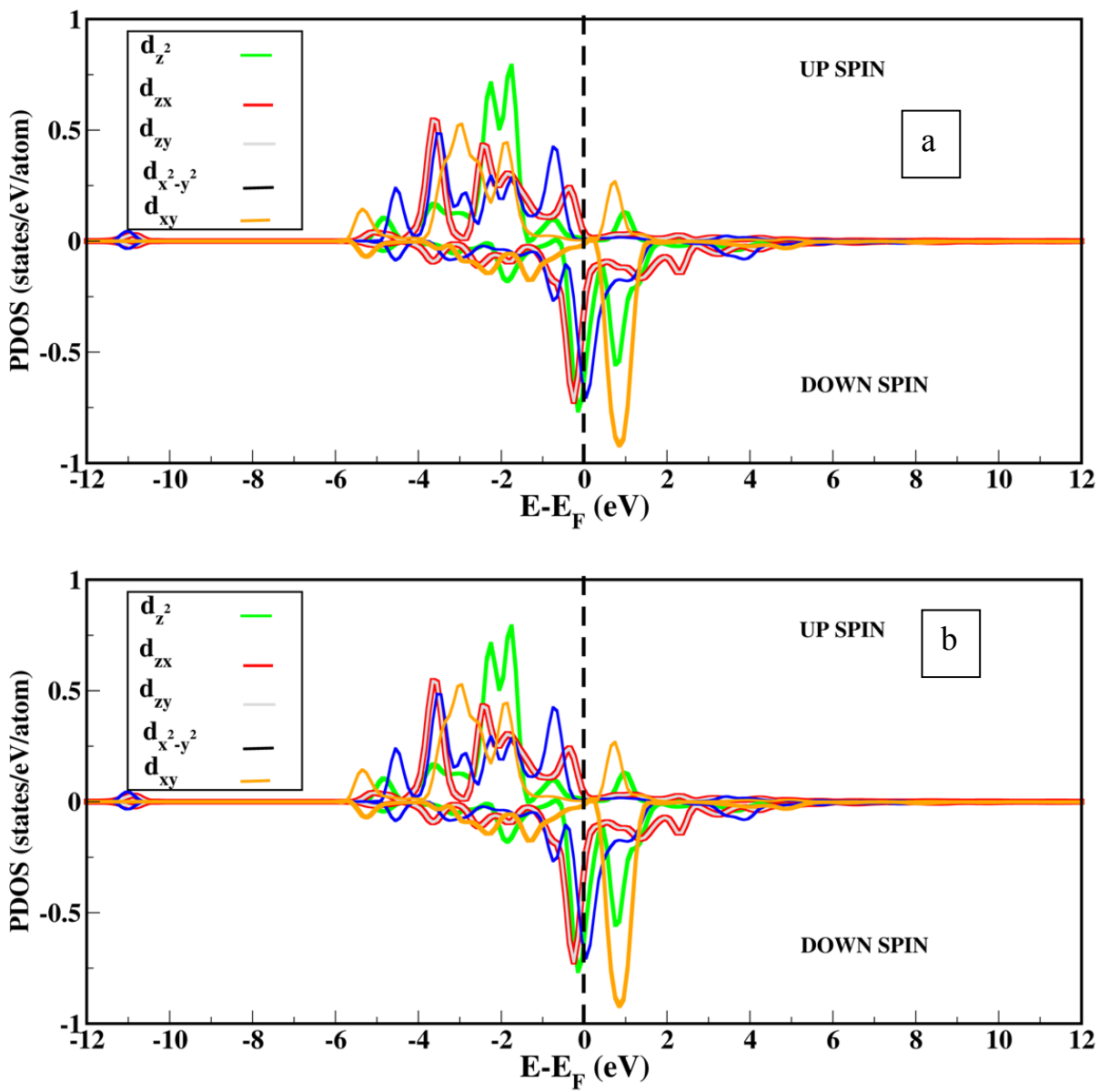

**Figure 11: contribution of different d orbitals :(a) type-A (b) type-B**

Clearly, most of the d orbitals are non-degenerate which gives clear indication of splitting of d orbital's. For type-A structure, spin-up component of $d_{zx}$ and $d_{zy}$ orbital are almost fully filled where as $d_{xy}$ component is partially filled. The splitting of d-orbital is caused by crystal field splitting with tetrahedral local coordination of Fe atoms. Though there is crystal field splitting of Fe '3d' orbitals due to tetrahedral coordination of Fe atoms, $d_{zx}$ and $d_{zy}$ still remain degenerate state in tetragonal symmetry.

## 2.4 Conclusions

We have systematically investigated the electronic structure of the superconducting system $CaFe_{1-x}Pt_xAs_2$ considering both types of existing crystal structures. Our calculated lattice parameters are in good agreement with experimental value. We find anti ferromagnetic stripe phase (AFM2) to be the ground state among all the other possible spin configurations which are in consistent with various iron-pnictide studies. Due to anti-ferromagnetic magnetic ordering (AFM2), we find anisotropic deformation in the crystal structure. We see the full metallic behavior of the parent compound which is consistent with experimental observation.[8] We find that there is significant overlap between Fe and As orbitals leading to hybridization. The Fe '3d' orbital's contribution is dominant at the Fermi level along with small contribution from As atomic orbitals. Hence Fe-As block plays the major role in the formation of superconductivity. Our calculated band diagram is highly dispersive and bands crossing the Fermi level give rise to electron and hole pocket in the Fermi surface. In the Fermi surface the dispersion along z-direction is quite low, which is quite unusual, interesting and novel in iron-pnictide studies. We observe the hole pocket at the Z point and electron pocket at the M point in the BZ which exclude the possibility of $s_{+-}$ symmetry of the superconducting order parameter. This finding is well suited with experimental predictions. The electronic structure of both the configurations (type-A and type-B) is similar. As electronic structure is closely related to the critical temperature ($T_C$), this finding also explains similar $T_C$ values obtained from experiments.